\documentclass[12pt,letterpaper]{JHEP3}

\newcommand{\sech}{\mathop{\mathrm{sech}}\nolimits}
\newcommand{\diag}{\mathop{\mathrm{diag}}\nolimits}
\newcommand{\halfp}{{\textstyle{\frac{p}{2}}}}
\newcommand{\quarterp}{{\textstyle{\frac{p}{4}}}}

\preprint{BROWN-HET-1479}

\title{\Large Semiclassical Quantization of the Giant Magnon}

\author{Georgios Papathanasiou and Marcus Spradlin\\
Department of Physics\\
Brown University, Providence, Rhode Island 02912, USA}

\abstract{
Solitons in field theory provide a window into regimes not
directly accessible by the fundamental perturbative degrees of freedom.
Motivated by interest in the worldsheet S-matrix of
string theory in $AdS_5 \times S^5$ in the limit of infinite
worldsheet volume we consider the semiclassical quantization of
a particular soliton of this theory:
the Hofman-Maldacena `giant magnon' spinning string.
We obtain explicit formulas for the complete spectrum of bosonic
and fermionic fluctuations around the giant magnon.
As an application of these results we confirm that the one-loop
correction to the classical energy vanishes as expected.
}

\begin{document}

\section{Introduction}

It is difficult to overstate the important role that solitons play in
quantum field theory, especially in integrable theories.
Because of their particle-like nature their dynamics can be effectively
captured by a small (in particular, finite) number of collective
degrees of freedom.
Quantization of these collective
coordinates
following the work of Gervais, Jevicki and
Sakita~\cite{Gervais:1974dc,Gervais:1975pa,Gervais:1975yg,Gervais:1976wr}
provides a systematic framework for studying the theory in regimes
not directly accessible by perturbation theory involving the
original degrees of freedom.

An interesting soliton to emerge in recent studies
of the AdS/CFT correspondence is the giant magnon of Hofman
and Maldacena~\cite{Hofman:2006xt}.  This is a soliton of the
integrable~\cite{Mandal:2002fs, Bena:2003wd}
$AdS_5 \times S^5$  worldsheet sigma-model~\cite{Metsaev:1998it}
(actually living
inside an ${\mathbb{R}} \times S^2$ subspace) whose image
in spacetime is a stretched string that is pointlike in $AdS_5$
and rotates uniformly around an axis of an $S^2 \subset S^5$.  Its name
derives from the fact that it is the dual description of an elementary
excitation (magnon)
in the spin chain description~\cite{Minahan:2002ve,
Beisert:2003yb,Beisert:2004hm,Arutyunov:2004vx,Staudacher:2004tk,
Beisert:2005fw,Beisert:2005tm}
of the dual ${\mathcal{N}} = 4$ gauge theory.

String theory on ${\mathbb{R}} \times S^2$ is classically
equivalent to sine-Gordon theory~\cite{Pohlmeyer:1975nb}
and the giant magnon is the image of the sine-Gordon soliton under
this equivalence.  Classical aspects of the giant magnon,
such as the phase shift for magnon scattering~\cite{Hofman:2006xt}, can
be understood in this way.  However string theory on $AdS_5 \times S^5$
is much more than just the reduced ${\mathbb{R}} \times S^2$ sigma-model,
having additional bosonic, fermionic and (in conformal gauge) ghost degrees
of freedom.
Correspondingly the giant magnon has a richer set of collective
coordinates (${\mathbb{R}} \times S^3$, together with eight
fermionic zero modes~\cite{Minahan:2007gf} which upon quantization
lead to 16 polarization states)
than the sine-Gordon soliton (just
the soliton's position ${\mathbb{R}}$).
The dressing method~\cite{Zakharov:1973pp}
(or the B\"acklund transformation~\cite{Mikhailov:2005zd})
can be used to construct
classical solutions~\cite{Spradlin:2006wk,Kalousios:2006xy}
describing arbitrary scattering and bound states (both
BPS~\cite{Dorey:2006dq,Chen:2006ge} and non-BPS~\cite{Hofman:2006xt})
of magnons.
Some of the latter have played an important role recently~\cite{Dorey:2007xn}
in elucidating the nature of double poles~\cite{Chen:2006gq,
Roiban:2006gs} in the
magnon S-matrix~\cite{Beisert:2005tm,Hernandez:2006tk,
Beisert:2006ib,Beisert:2006ez}.
Other related aspects of giant magnons
and the S-matrix have been studied recently
in~\cite{Eden:2006rx,Arutyunov:2006iu,
Arutyunov:2006gs,Minahan:2006bd,Chu:2006ae,Plefka:2006ze,Okamura:2006zv,
Arutyunov:2006ak,Chen:2006gp,
Arutyunov:2006yd,
Astolfi:2007uz,Vicedo:2007rp,Kluson:2007qu}.

The study of classical spinning strings~\cite{Gubser:2002tv}
and especially their semiclassical quantization (see in
particular~\cite{Frolov:2002av,Frolov:2003tu}
and the many additional references given
below) has been incredibly fertile ground for quantitative studies
of AdS/CFT.
Typically the semiclassical analysis is done at the level of the lagrangian,
by expanding the action to quadratic order in
small fluctuations around the classical
background and looking for appropriate field redefinitions
to decouple various independent modes from each other.
One aspect of the giant magnon that sets it apart from much of
this extensive literature is the fact that it is not at rest on
the worldsheet (in conformal gauge, which is by far the most natural
gauge to use in this case).  This fact makes it more natural to
carry out the semiclassical analysis not at the level of fields in
the lagrangian but rather by working out an explicit basis of
eigenmodes
(i.e., asymptotically plane-wave solutions of the linearized equations
of motion around the soliton background).
The present work was motivated in part by~\cite{Minahan:2007gf}
where Minahan explicitly constructed the fermionic zero modes of
the giant magnon but left analysis of the non-zero mode fluctuations
tantalizingly open in his equation (4.1).
Our work benefits in particular from using the same convenient
form of $\kappa$-symmetry fixing employed in~\cite{Minahan:2007gf}.

The outline of this paper is as follows.
In section 2 we review the giant magnon solution
and display explicit solutions for small bosonic fluctuations
around the classical background.  This part requires relatively little
work since the dressing method can be used to algebraically construct
a complete set of small fluctuations around any $N$-soliton
configuration.  We do not know of a similar construction
for fermionic fluctuations, which we therefore carry out `by hand' in section 3.
In fact the fermionic analysis turns out to be much
simpler than we had any right to expect.
Some steps of the calculation
fall together in an almost miraculous way, encouraging us to speculate
that properly understanding the structure should make it feasible to
construct fermionic fluctuations around more complicated multi-soliton
backgrounds.

Finally in section 4 we use our results to read off the stability angles
which appear in the evaluation of the 1-loop functional determinant in
the soliton background,
following the method pioneered by
Dashen, Hasslacher and Neveu~\cite{Dashen:1975hd}.
We show that the first
${\cal{O}}(1/\sqrt{\lambda})$
correction to the classical `energy' $\Delta - J$ of the giant magnon
vanishes.
Although this particular result is in accord with the
expectation based on the $SU(2|2) \ltimes {\mathbb{R}}^2$
superalgebra~\cite{Beisert:2005tm,Hofman:2006xt},
continued study of quantized giant magnons
should provide an approach to probing the $AdS_5 \times S^5$ worldsheet
S-matrix that complements studying it via
the scattering of
the fundamental worldsheet degrees of freedom~\cite{Klose:2006zd}.
The latter approach has proven to be surprisingly
successful at one loop~\cite{Klose:2007wq}
in a certain limit introduced by Maldacena and
Swanson~\cite{Maldacena:2006rv}.
Unfortunately taking this limit seems to complicate the analysis
of giant magnons considerably, even at the classical level.

\section{Bosonic Sector}

We write the bosonic part of the $AdS{}_5 \times S^5$ sigma-model
in the form~\cite{Arutyunov:2003za}
\begin{equation}
\label{eq:bosonicaction}
S =
- {\sqrt{\lambda} \over 2 \pi} \int dt\,dx\  \Big[
\eta^{ab}
\partial_a Y^\mu \partial_b Y_\mu
+ \widetilde{\Lambda}
(Y^2 + 1)\Big]
+ \Big[\eta^{ab}
\partial_a X^i \partial_b X_i + \Lambda (X^2 - 1)\Big],
\end{equation}
with $\eta^{ab} = \diag(-1,+1)$
and $\lambda$ identified with the `t Hooft coupling in the dual gauge theory.
Here $Y^\mu$ are coordinates on
$\mathbb{R}^{4,2}$, $X^i$ are coordinates on
$\mathbb{R}^6$, and the Lagrange multipliers $\widetilde{\Lambda}$
and $\Lambda$ enforce the embedding constraints
\begin{equation}
\label{eq:embedding}
Y^2 = -1\,, \qquad X^2 = +1
\end{equation}
onto $AdS{}_5$ and $S^5$ respectively.
The sigma-model equations of motion
\begin{equation}
\label{eq:bosoniceom}
(\partial^2 - \widetilde{\Lambda}) Y = (\partial^2 - \Lambda) X = 0
\end{equation}
are to be supplemented in string theory by the Virasoro constraints
\begin{eqnarray}
\label{eq:virasoro}
T_{00} &=& \frac{1}{2} \left[ (\partial_t Y)^2 + (\partial_x Y)^2 +
(\partial_t X)^2 + (\partial_x X)^2
\right] = 0\,,\cr
T_{01} &=& \partial_t Y^\mu \partial_x Y_\mu
+ \partial_t X^i \partial_x X_i = 0\,.
\end{eqnarray}
{}From~(\ref{eq:embedding}) and~(\ref{eq:bosoniceom})
it follows
that the classical values of the Lagrange multipliers are
\begin{equation}
\label{eq:lagrange}
\widetilde{\Lambda} = - Y^\mu \partial^2 Y_\mu\,, \qquad
\Lambda = + X^i \partial^2 X_i\,.
\end{equation}

\subsection{The giant magnon}

For the rest of the paper we will be interested in a particular
classical solution of the
equations~(\ref{eq:bosoniceom}) and~(\ref{eq:virasoro}), namely
the giant magnon of~\cite{Hofman:2006xt}.
This is an open string
which is pointlike
in $AdS_5$ and extended along
the $S^5$, with endpoints moving at the speed of light along
an equator of the $S^5$, which we choose to lie in the
$X^5$--$X^6$ plane.
We introduce the notation $\vec{X}$ to refer
to the four transverse coordinates $X^1,\ldots,X^4$.
A physical closed string
is obtained by attaching two or more giant magnons to each other
end to end. For
our purposes we can ignore this step because the individual pieces of
string
do not talk to each other.

The giant magnon is characterized by a `momentum' $p \in [0,2 \pi)$
and by the choice of a point on $S^3$, i.e.~by a
a four-component unit vector $\vec{n}$, which specifies
the polarization of the magnon in the transverse directions.
Choosing $Y^0$ and $Y^5$ to be the two timelike directions
in $\mathbb{R}^{4,2}$, the giant magnon solution
in our coordinates takes the form
\begin{eqnarray}
\label{eq:giantmagnon}
Y^0 + i Y^5 &=& e^{i t}\,,\cr
\vec{X} &=& \vec{n}\,\sin\halfp\sech u\,, \cr
Z \equiv
X^5 + i X^6 &=& e^{i t} \left[ \cos \halfp + i \sin \halfp\tanh u \right],
\end{eqnarray}
where
\begin{equation}
\label{eq:udef}
u = (x - t \cos\halfp) \csc \halfp\,.
\end{equation}
The Lagrange multipliers~(\ref{eq:lagrange}) are classically equal to
\begin{equation}
\label{eq:lagrangevalue}
\widetilde{\Lambda} = 1\,, \qquad
\Lambda = 1 - 2 \sech^2 u\,.
\end{equation}
It is evident that the solution~(\ref{eq:giantmagnon})
describes a localized wave (i.e., a soliton)
polarized in the $\vec{n}$ direction
which travels along the worldsheet with velocity $\cos \halfp$.
The quantity $p$ is not a Noether charge of
the action~(\ref{eq:bosoniceom}) (the charge associated with $x$
translations
is identically zero due to the Virasoro constraints).  Rather it is
a parameter of the soliton~(\ref{eq:giantmagnon})
which specifies
the difference in longitude between the two endpoints of the string.
The quantity $e^{ip}-1$
appears as a central charge in the supersymmetry algebra of the
giant magnon~\cite{Beisert:2005tm,
Hofman:2006xt,Arutyunov:2006ak},
and while individual magnons can have arbitrary $p$,
physical closed string states
have total momentum $p \in 2 \pi {\mathbb{Z}}$ so that
$e^{i p} - 1 = 0$.

The giant magnon has four bosonic collective coordinates
(zero modes) parameterizing
${\mathbb{R}} \times S^3$.  The $S^3$ coordinate is just the polarization
vector $\vec{n}$ we already encountered.  The ${\mathbb{R}}$
coordinate, which we may call $q$, denotes the position
of the magnon on the worldsheet; this collective coordinate may be
introduced by replacing $x \to x - q$ in~(\ref{eq:udef}).
In addition the giant magnon is a BPS state
and there are eight fermionic collective
coordinates which have been explicitly constructed in~\cite{Minahan:2007gf}.

\subsection{$AdS_5$ fluctuation spectrum}

We now turn to the analysis of the spectrum of fluctuations around
the giant magnon solution~(\ref{eq:giantmagnon}), beginning with
the $AdS_5$ bosons, where we find one massless and four massive
fluctuations.  The reader is unlikely to be surprised by this result,
but it is instructive to treat the analysis carefully in order to set
the stage for the following sections.

The equation of motion for a fluctuation $\delta Y^\mu$
around the giant magnon solution, after eliminating the Lagrange
multiplier, is
\begin{equation}
\label{eq:adsfluct}
(\partial^2 - 1) \delta Y^\mu +
Y^\mu Y_\nu \partial^2 \delta Y^\nu = 0\,,
\end{equation}
subject to the constraint
\begin{equation}
\label{eq:adsfluctconstraint}
Y_\mu\,\delta Y^\mu = 0\,.
\end{equation}

The massless fluctuation
is exhibited by solving~(\ref{eq:adsfluctconstraint})
with the ansatz
\begin{equation}
\delta Y^0 = - f \sin t\,, \qquad \delta Y^5 = + f \cos t\,.
\end{equation}
Substituting this into~(\ref{eq:adsfluct}) we find that $f$
satisfies the free wave equation
\begin{equation}
\label{eq:massless}
\partial^2 f = 0\,.
\end{equation}
In a proper quantization of string theory on $AdS_5 \times S^5$
the sigma-model action~(\ref{eq:bosonicaction}) would be supplemented
by ghosts which would cancel the mode~(\ref{eq:massless}) together
with a similar $S^5$ mode that we will find below.
These are analogous to longitudinal fluctuations in light-cone gauge, and
it is sufficient for the purpose of our semiclassical analysis to
simply omit them~\cite{Frolov:2002av,Park:2005ji}.

The remaining
four fluctuations $\delta \vec{Y}$ in the transverse spatial
directions of $AdS_5$ satisfy~(\ref{eq:adsfluctconstraint})
automatically, and moreover for these modes the second
term in~(\ref{eq:adsfluct}) vanishes, giving
four free bosons with mass $m^2 = 1$.
Note that these modes preserve the Virasoro
constraints~(\ref{eq:virasoro}) to first order in the fluctuation.

\subsection{$S^5$ fluctuation spectrum}

The equation of motion for an $S^5$ fluctuation $\delta X^i$
around~(\ref{eq:giantmagnon}) is
\begin{equation}
\label{eq:spherefluct}
(\partial^2 - 1 + 2 \sech^2 u)  \delta X^i - X^i X_j \partial^2 \delta X^j = 0
\end{equation}
subject to
\begin{equation}
\label{eq:sphereconstraint}
X_i\,\delta X^i = 0\,.
\end{equation}
These equations admit two different classes of solutions.

The first class of solutions are the zero modes which have
been discussed in~\cite{Minahan:2007gf}
and are in one-to-one correspondence with the collective
coordinates.
The choice of polarization $\vec{n}$ breaks the $SO(4)$ symmetry of
the transverse directions, leading to the three independent zero modes
\begin{equation}
\delta \vec{X} = \vec{m} \sech u\,, \qquad \vec{m} \cdot \vec{n} = 0\,.
\end{equation}
Moreover the presence of the magnon breaks the $x$-translation
symmetry, leading to the zero mode
\begin{eqnarray}
\delta \vec{X} &=& - \vec{n} \sech u \tanh u\,, \cr
\delta Z &=& i e^{i t} \sech^2 u\,.
\end{eqnarray}
These particular solutions will not play any further role in our analysis.

Instead of these normalizable zero modes we are
interested
in plane wave fluctuations of the form
\begin{equation}
e^{i k u - i \omega v} f(u)\,,
\end{equation}
where $f(u)$ is some profile and
we have introduced
\begin{equation}
\label{eq:vdef}
v = (t - x \cos \halfp) \csc \halfp
\end{equation}
as the time variable in the magnon's `rest frame'
complementing $u$ as defined in~(\ref{eq:udef}).
The variables $u$ and $v$ resemble variables
appropriate for a relativistic
system boosted to velocity $\cos \halfp$, but
it should be kept in mind that
the giant magnon
solution~(\ref{eq:giantmagnon}) is not
relativistically invariant~\cite{Hofman:2006xt}.
In particular, recall that
the parameter $p$ represents the physical separation in
longitude of the
two endpoints of the string on $S^2$.
The appearance of these apparently boosted variables emerges from
the fact that string theory on ${\mathbb{R}} \times S^2$
is classically equivalent to relativistic sine-Gordon theory.

One of the plane-wave solutions of~(\ref{eq:spherefluct}) is
massless.  A basis for the positive energy
excitations of this mode
is given by
\begin{eqnarray}
\delta \vec{X} &=&  e^{i k u - i |k| v}
 \vec{n}\,
 (k + |k| \cos \halfp) \sech u \tanh u\,,
\cr
\delta Z &=&  -i e^{i k u - i |k| v}
e^{+i t} \left[ k - |k| \sinh u \sinh(u+i
\halfp) \right]\sech^2 u\,,\cr
\delta \overline{Z} &=& +i e^{i k u - i |k| v}
e^{-i t}  \left[
 k - |k| \sinh u \sinh(u - i \halfp)  \right]\sech^2 u\,,
\end{eqnarray}
where $k \in {\mathbb{R}}$ is the wavenumber.
In string theory we will omit this mode as discussed
beneath~(\ref{eq:massless}).

The remaining four physical fluctuations are characterized by
a polarization vector $\vec{m}$ in the transverse ${\mathbb{R}}^4$.
The corresponding positive frequency excitations are
\begin{eqnarray}
\label{eq:spherefluctsolution}
\delta \vec{X} &=&
e^{i k u - i \omega v}
\left[
\vec{m}  (k + i \tanh u)
- \vec{n} (n \cdot m) \left(k + \omega \cos \halfp\right)
\sech^2 u
\right],\cr
\delta Z &=& - i e^{i k u - i \omega v} e^{+i t} {(n \cdot m)}
\left[
 k \sinh u +   \omega \sinh(u + i \halfp)
+ i \cosh u
\right] \sech^2 u\,, \cr
\delta \overline{Z} &=& +i e^{i k u - i \omega v}
e^{-i t} (n \cdot m)
\left[
 k \sinh u +  \omega \sinh(u - i \halfp)
+ i \cosh u
\right]\sech^2 u\,,
\end{eqnarray}
where
\begin{equation}
\omega = + \sqrt{k^2 + 1}
\end{equation}
indicating that the effective mass of these modes is
$m^2 = 1$.
Indeed, very far from the soliton core ($|u| \gg 1$) the fluctuation
$\delta \vec{X}$ satisfies the free massive wave equation and
its profile becomes constant.
Note that the three fluctuations orthogonal to the original
giant magnon ($n \cdot m = 0$) take the very simple form
\begin{equation}
\delta \vec{X} = e^{i k u - i \omega v} \vec{m} (k + i \tanh u)\,.
\end{equation}

Although it is straightforward to verify that~(\ref{eq:spherefluctsolution})
satisfies~(\ref{eq:spherefluct}) and~(\ref{eq:sphereconstraint}) as
claimed, a comment is in order regarding how we obtained
this rather complicated formula.
In fact it is well known~\cite{Dashen:1975hd}
(see~\cite{Gervais:1976wr,deVega:1982sh} for further examples)
that a complete basis of bosonic fluctuations
around any
$N$-soliton solution in a classically integrable theory can be constructed
algebraically, for example
by the dressing method \cite{Zakharov:1973pp} or equivalently the
B\"acklund transformation (see in particular \cite{Mikhailov:2005zd} for
a nice discussion in this context).
One begins with the desired background
solution and then superposes on top of it a soliton-antisoliton bound
state (i.e., a breather).   Expanding the resulting solution to first order
in the (complex)
momentum of the breather yields the desired fluctuation of the
background solution.  Unfortunately we know of no similarly simple
way to determine the fermionic fluctuations in the following section.

\section{Fermionic Sector}

In this section we obtain explicit formulas for the complete
spectrum of fermionic
fluctuations around the giant magnon solution~(\ref{eq:giantmagnon}).
The classical background has $\vartheta = 0$ so we can use
the variable $\vartheta$ itself to denote the fluctuation, rather
than the more cumbersome alternative $\delta \vartheta$.
The action for the fermionic fluctuations around a general classical string
solution ${\displaystyle{{\mathrm{X}}^\mu}}(t,x)$ is~\cite{Metsaev:1998it}
\begin{equation}
\label{eq:fermionaction}
S = {\sqrt{\lambda} \over 2 \pi} \int dt\,dx\ \Big[
i (\eta^{ab} \delta^{IJ} - \epsilon^{ab} s^{IJ})
\bar{\vartheta}^I \rho_a D_b \vartheta^J + {\cal O}(\vartheta^4)\Big].
\end{equation}
The quartic terms will play no role in our linearized analysis of
fluctuations around $\vartheta = 0$.
Here $\vartheta^I$ are two 10-dimensional Majorana-Weyl spinors,
$s^{IJ} = \diag(+1,-1)$,
\begin{equation}
\rho_a = \Gamma_A E^A_\mu({\mathrm{X}}) \partial_a {\mathrm{X}}^\mu
\end{equation}
are projections of the ten-dimensional Dirac matrices involving
the $AdS_5 \times S^5$ vielbein $E^A_\mu$ and the bosonic coordinates
${\displaystyle{{\mathrm{X}}^\mu}}$ of the classical solution.
Note that $\mu$ is now a curved space index denoting all
of the $AdS_5 \times S^5$ directions, rather than the flat space
index in the embedding space ${\mathbb{R}}^{4,2}$ of the $AdS_5$ part
only, that we used in the previous section.
The covariant derivative appearing in~(\ref{eq:fermionaction})
is given by
\begin{equation}
D_a \vartheta^I = \left(
\delta^{IJ} (\partial_a +  {\textstyle{\frac{1}{4}}}\omega_\mu^{AB}
\partial_a {\displaystyle{{\mathrm{X}}^\mu}}
\Gamma_{AB})
- {\textstyle{\frac{i}{2}}} \epsilon^{IJ} \Gamma_* \rho_a
\right) \vartheta^J
\end{equation}
in terms of the usual spin connection $\omega_\mu^{AB}$ and the matrix
\begin{equation}
\Gamma_* = i \Gamma_{01234}
\end{equation}
satisfying $\Gamma_*^2 = 1$.
Whenever necessary, we use for concreteness a Majorana representation
of imaginary Dirac matrices with
$\Gamma_0$ symmetric and the other nine matrices
antisymmetric.

\subsection{The fluctuation equations}

The linearized
equations of motion following from~(\ref{eq:fermionaction}) are
\begin{eqnarray}
\label{eq:fermioneom}
(\rho_0 - \rho_1)(D_0 + D_1) \vartheta^1 &=& 0\,,\cr
(\rho_0 + \rho_1)(D_0 - D_1) \vartheta^2 &=& 0\,.
\end{eqnarray}
The full Green-Schwarz action for $AdS_5 \times S^5$, including
both~(\ref{eq:bosonicaction}) and~(\ref{eq:fermionaction}),
possesses a local fermionic symmetry ($\kappa$-symmetry) under
which the physical solutions must be gauge-fixed.
The zero mode solutions of~(\ref{eq:fermioneom}) were recently
constructed by Minahan~\cite{Minahan:2007gf}, who noted that
it is simplest to first find solutions without worrying about
$\kappa$-symmetry and impose a projection only at the end of the
calculation.
We will find that the non-zero mode solutions of~(\ref{eq:fermioneom})
enjoy this simple $\kappa$-fixing as well.

We begin by following~\cite{Minahan:2007gf} to process~(\ref{eq:fermioneom})
into a more convenient form.
We change variables from $(x,t)$ to
the coordinates
\begin{eqnarray}
u &=& (x - t \cos \halfp) \csc \halfp\,,\cr
v &=& (t - x \cos \halfp) \csc \halfp
\end{eqnarray}
introduced in~(\ref{eq:udef}) and~(\ref{eq:vdef})
above (note that these were denoted respectively by $(x,\xi)$
in~\cite{Minahan:2007gf}).
The equations for $\vartheta^I$ can be decoupled and written as
\begin{equation}
\label{eq:minahanone}
i (\rho_0 - \rho_1) \left[\frac{1}{\tanh u}
({\cal D} - \partial_v) \frac{1}{\tanh u} ({\cal D}
+ \partial_v) \vartheta^1 - \vartheta^1\right] = 0\,,
\end{equation}
\begin{equation}
\label{eq:minahantwo}
i (\rho_0 + \rho_1) \left[
\frac{1}{\tanh u}(\widetilde{\cal D} - \partial_v) \frac{1}{\tanh u}
(\widetilde{\cal D} +
\partial_v) \vartheta^2 - \vartheta^2\right] = 0\,,
\end{equation}
where
\begin{eqnarray}
\label{eq:fermionstep}
{\cal D} &=& \partial_u +
\frac{1}{2} G \Gamma_{\phi \theta}, \qquad
G = +\frac{1}{ \cosh u}\left(
1 + \frac{\cos \halfp}{\tanh^2 u + \cos^2 \halfp \sech^2 u}
\right),\cr
\widetilde{\cal D} &=& \partial_u
+ \frac{1}{2} \widetilde{G} \Gamma_{\phi \theta}, \qquad
\widetilde{G} =  - \frac{1}{ \cosh u} \left(
1 - \frac{\cos \halfp}{\tanh^2 u + \cos^2 \halfp \sech^2 u} \right).
\end{eqnarray}
The subscripts on $\Gamma_{\phi \theta}$ refer
to the usual angular coordinates
$(\theta,\phi)$ parameterizing the $S^2 \subset S^5$ on which
the giant magnon~(\ref{eq:giantmagnon}) lives
(i.e.~$\vec{X} = \vec{n} \cos \theta$, $Z = e^{i \phi} \sin \theta$).
The matrix $\Gamma_{\phi \theta}$ satisfies $(\Gamma_{\phi \theta})^2 = -1$,
and in fact this is all we need to know about it.

The matrices $\rho_0 \pm \rho_1$
each have half the maximal rank and commute with the differential
operators ${\cal D}$ and
$\widetilde{\cal D}$.
Following~\cite{Minahan:2007gf},
we can therefore
first solve~(\ref{eq:minahanone}) and~(\ref{eq:minahantwo})
for $\vartheta^I$ and then treat the projected spinors
\begin{equation}
\label{eq:projection}
\Psi^1 = i (\rho_0 - \rho_1) \vartheta^1\,, \qquad
\Psi^2 = i (\rho_0 + \rho_1) \vartheta^2\,
\end{equation}
as the fields fixed under
$\kappa$-symmetry.
For later reference we note equation (2.16) from~\cite{Minahan:2007gf},
which in our notation reads
\begin{equation}
\label{eq:useful}
\tan \quarterp ({\cal D} + \partial_v) \Psi^1 - {\textstyle{\frac{i}{2}}}
\Gamma_* (\bar{\rho}_0 - \rho_0) \Psi^2 = 0\,,
\end{equation}
where $\bar{\rho}_0 = - \rho_0^\dagger = \Gamma_* \rho_0 \Gamma_*$.
Since the matrix $(\bar{\rho}_0 - \rho_0)$
is nonsingular
this equation determines $\Psi^2$ completely once $\Psi^1$ is known.

\subsection{Solving the equation for $\vartheta$}

Here we present the technical details involved in solving
the equations~(\ref{eq:minahanone}) and~(\ref{eq:minahantwo}).
The reader interested only in the final answers may turn to the
next subsection.
Since the two equations have
nearly identical form we begin by focusing on the
unprojected version of the first equation
\begin{equation}
\label{eq:minahan}
\frac{1}{\tanh u}
({\cal D} - \partial_v) \frac{1}{\tanh u} ({\cal D}
+ \partial_v) \vartheta^1 - \vartheta^1 = 0\,.
\end{equation}
We proceed
by transforming the second order equation~(\ref{eq:minahan})
into a system of
two coupled first ordered equations
\begin{eqnarray}
\label{eq:one}
\frac{1}{\tanh u} \left( {\cal D}
+ \partial_v \right) \vartheta^1 &=& \widetilde{\vartheta}^1\,,\cr
\frac{1}{\tanh u} \left( {\cal D}
- \partial_v \right) \widetilde{\vartheta}^1 &=& \vartheta^1\,
\end{eqnarray}
involving $\vartheta^1$ and a new field $\widetilde{\vartheta}^1$
(which in fact
is clearly related to $\vartheta^2$ thanks to~(\ref{eq:useful})).
In what follows we employ the matrix notation
\begin{equation}
\Theta = \pmatrix{
\vartheta^1 \cr
\widetilde{\vartheta}^1 }.
\end{equation}

The zero mode
solutions satisfying $\partial_v \vartheta^I=0$
were constructed by Minahan in~\cite{Minahan:2007gf}.
For the non-zero modes we make a Fourier ansatz for the $v$
dependence
\begin{equation}
\label{eq:fourier}
\Theta(v,u) = e^{-i \omega v} \Theta(u)\,.
\end{equation}
Since $(\Gamma_{\phi \theta})^2 = - 1$
we can decompose
$\Theta$ into eigenspinors as
\begin{equation}
\Theta = \Theta_+ + \Theta_-, \qquad \Gamma_{\phi \theta}
\Theta_\pm = \pm i \Theta_\pm\,.
\end{equation}
Then using ${\cal D} = \partial_u + \frac{1}{2} G \Gamma_{\phi \theta}$
we see that~(\ref{eq:one}) can be rearranged into the matrix equation
\begin{equation}
\label{eq:ddd}
(\partial_u - A_\pm) \Theta_\pm = 0, \qquad
A_\pm = \pmatrix{
i (\omega \mp \frac{G}{2}) & \tanh u \cr
\tanh u & i (\omega \pm \frac{G}{2} )}.
\end{equation}
Our strategy is to find an invertible transformation
\begin{equation}
\label{eq:ccd}
\Theta_{\pm}
\to \Theta'{}_\pm = S \Theta_\pm
\end{equation}
that diagonalizes~(\ref{eq:ddd}),
solve the system of equations for $\Theta'{}_{\pm}$
and then transform
back to find the result for
$\Theta_{\pm}$.

In the transformed variables~(\ref{eq:ddd}) becomes
\begin{equation}
(\partial_u - A'_\pm) \Theta'{}_\pm = 0, \qquad
A'_\pm = (\partial_u S + S A_\pm) S^{-1}\,.
\end{equation}
Parameterizing
\begin{equation}
S = \pmatrix{ a(u) & b(u) \cr
c(u) & d(u) }
\end{equation}
we find that $A'_{\pm}$ are both diagonal if these entries satisfy
\begin{eqnarray}
\label{eq:hhh}
a\, b' - b\, a' + (a^2 - b^2) \tanh u - 2 i \omega a b &=& 0\,, \cr
c\, d' - d\, c' + (c^2 - d^2) \tanh u - 2 i \omega c d &=& 0\,,
\end{eqnarray}
where the prime denotes $\partial/\partial u$.
The fact that these
equations are the same for both
$A_+$ and $A_-$ means that the equations for
both $\Theta_+$ and $\Theta_-$ can be
simultaneously diagonalized
by the same matrix $S$.

We notice that we have two differential equations for four unknown functions,
which simply signifies that given a transformation matrix $S$
which diagonalizes~(\ref{eq:ddd})
then the same will be true if we multiply $S$ by an arbitrary
diagonal matrix.
The final result of our
calculation must of course be independent of these two extra degrees
of freedom, so we may choose to fix them by imposing the
convenient constraints
\begin{eqnarray}
\label{eq:iii}
a' + b \tanh u &=& 0\,, \cr
c' + d \tanh u &=& 0\,.
\end{eqnarray}
This choice considerably simplifies $S$ and $A'_\pm$, which
we find now takes the
form
\begin{equation}
\label{eq:iij}
A'_\pm = \pmatrix{i (\omega \mp \frac{G}{2} ) & 0 \cr
0 & i (\omega \mp \frac{G}{2})},
\end{equation}
completely independent of
the transformation matrix elements $a,b,c,d$.
Even more remarkably, the resulting diagonal equation
$(\partial_u - A'_\pm) \Theta'{}_\pm = 0$ is solved by an arbitrary constant
spinor times $e^{i \omega u} e^{\pm i \chi}$, where $\chi$
is precisely the same function
which appeared in the zero mode solutions of~\cite{Minahan:2007gf} (and
shown in~(\ref{eq:chidef}) below).

All that remains is to determine the elements of the transformation
matrix $S$.  With the choice~(\ref{eq:iii}),~(\ref{eq:hhh}) becomes
\begin{eqnarray}
\label{eq:jjj}
b' + a \tanh u - 2 i \omega b &=& 0\,, \cr
d' + c \tanh u - 2 i \omega d &=& 0\,.
\end{eqnarray}
The general solution to the coupled system~(\ref{eq:iii}),~(\ref{eq:jjj}) is
\begin{equation}
S = \pmatrix{ a(u) & b(u) \cr
c(u) & d(u)} =
S_0
\pmatrix{
- |\omega + k|\, a_1(u) & e^{2 i \omega u} b_1(u) \cr
- |\omega - k|\, a_2(u) & e^{2 i \omega u} b_2(u)},
\end{equation}
where $S_0$ is an arbitrary constant matrix
(though we want it to be invertible),
\begin{eqnarray}
\label{eq:bonetwo}
b_1(u) &=& e^{i (-\omega - k) u}
e^{\frac{i}{2} \left(  \tan^{-1} (\omega \sinh 2 u) +
\tan^{-1}(k \tanh 2 u) \right) } \sech u \sqrt{|\omega\cosh 2u - k|}\,,\cr
b_2(u) &=& e^{i (-\omega + k) u}
e^{\frac{i}{2} \left(  \tan^{-1} (\omega \sinh 2 u) -
\tan^{-1}(k \tanh 2 u) \right) } \sech u \sqrt{|\omega\cosh 2u + k|}\,,
\end{eqnarray}
the $a_i(u)$ are obtained
from the $b_i(u)$ by replacing $\omega \to -\omega$, and
we have introduced
\begin{equation}
k = + \sqrt{\omega^2 - 1}.
\end{equation}
We restrict our attention to $|\omega| \ge 1$ so that $k$ is real and
we get plane wave normalizable solutions.

Next we transform back to our initial spinors
$\Theta_\pm = S^{-1} \Theta'{}_\pm$.
We find that the inverse of $S$ is simply
\begin{equation}
\Theta_\pm=
S^{-1} \Theta'{}_\pm = - \frac{1}{4 \omega k}
\pmatrix{
b_2(u) & - b_1(u) \cr
e^{-2 i \omega u} |\omega - k|\, a_2(u) & - e^{-2 i \omega u} |\omega + k|
\, a_1(u)} S_0^{-1} \Theta'{}_\pm.
\end{equation}
Recalling that
$\Theta'{}_\pm$ was found to be given by $e^{i \omega u} e^{\pm i \chi}$
times
an arbitrary constant spinor it is clear that we can absorb the
constants of integration $S_0^{-1}$ into the choice of constant
spinor.
Then the most general solution for $\vartheta_\pm^1$ (the
top component of $\Theta_\pm$) is
of the form
\begin{equation}
\label{eq:first}
\vartheta_\pm^1 = e^{i \omega u} e^{\pm i \chi}
\left[ b_2(u) U^2_\pm + b_1(u) U_\pm^1 \right],
\end{equation}
where $U^i_\pm$ are arbitrary constant spinors satisfying
$\Gamma_{\phi \theta}  U_\pm^i = \pm i  U_\pm^i$.

However it is clear from~(\ref{eq:bonetwo}) that $b_1(u)$ and $b_2(u)$
differ
only by $k \leftrightarrow -k$.
Therefore it is more transparent to treat the wavenumber $k$, rather
than the frequency $\omega$, as the parameter of the solution.
Restoring the  $e^{-i \omega v}$ dependence
from~(\ref{eq:fourier}), we therefore conclude that the most general
solution to~(\ref{eq:minahan}) is
\begin{equation}
\label{eq:positive}
\vartheta^1 = \sech u
\sqrt{|\omega \cosh 2 u + k|}\,e^{i \alpha}
\left[ e^{+ i \chi} U_+ + e^{-i \chi} U_-\right],
\end{equation}
where we now allow $k \in (-\infty,+\infty)$ with
$\omega^2 = {k^2 + 1}$,
$U_\pm$ are arbitrary ($k$-dependent) Weyl spinors
satisfying $\Gamma_{\phi \theta} U_\pm =
\pm i U_\pm$, and
\begin{equation}
e^{i \alpha} = e^{ - i \omega v + i k u}
e^{\frac{i}{2} \left( +\tan^{-1} (\omega \sinh 2 u) -
\tan^{-1}(k \tanh 2 u) \right) }.
\end{equation}
Far away from the core of the magnon ($|u| \gg 1$) the
solution behaves like $e^{-i \omega v + i k u}$,
representing a free fermion of mass $m^2 = 1$.

Of course the second order equation~(\ref{eq:fermioneom}) admits
both positive frequency ($\omega>0$) and negative frequency ($\omega<0$)
solutions, and we have
been careful in writing~(\ref{eq:positive})
to allow both cases.  However
the two kinds of solutions
are clearly related to each other by complex conjugation (together
with relabeling $k \to -k$ and $U_\pm \to U^*_\mp$) so we do not
need to treat them separately.
In what follows we focus on the positive frequency solutions,
postponing discussion of the Majorana condition $(\Psi^I)^* =
\Psi^I$ until the end.

\subsection{The $\kappa$-fixed solutions}

It remains to impose the $\kappa$ symmetry projection~(\ref{eq:projection}).
After some elementary Dirac matrix algebra we find
the positive frequency mode
\begin{equation}
\label{eq:second}
\Psi^1 = i \sech u \sqrt{\omega \cosh 2 u + k}\, e^{i \alpha}
\left[ e^{+ i \chi} \Gamma_0 + e^{-i \chi} \Gamma_\phi \right]
(U_+
-\Gamma_0 \Gamma_\phi U_-)\,.
\end{equation}
Let us now address the counting of degrees of freedom.
Before imposing any chirality
condition, each $U_\pm$ has 16 (complex) components,
so it might appear that~(\ref{eq:second}) involves 32 independent
components.
However if we assemble $U_\pm$ into a single
32-component spinor $U = U_+ + U_-$ according to
\begin{equation}
U_\pm = \frac{1}{2 i} (i \pm \Gamma_{\phi \theta}) U\,,
\end{equation}
then the combination appearing in~(\ref{eq:second}) is
\begin{equation}
\label{eq:pdef}
(U_+ - \Gamma_0 \Gamma_\phi U_-) = {\cal P} U\,, \qquad
{\cal P} = {1 \over 2 i} \left[(i + \Gamma_{\phi \theta}) -
\Gamma_0 \Gamma_\phi (i - \Gamma_{\phi \theta}) \right].
\end{equation}
It is not hard to check that ${\cal P}$ is a projection matrix with
half maximal rank, so in fact~(\ref{eq:second}) only involves the 16
components of $U$ which are not killed by ${\cal P}$.
Moreover ${\cal P}$ commutes with the chirality operator $\Gamma^{11}$ and
has eight nontrivial eigenspinors of each chirality.  Therefore
it is consistent to impose a Weyl condition
on $U$ further reducing the number of components by half
to just eight.

To summarize, we find the general $\kappa$-fixed positive frequency
modes
\begin{equation}
\label{eq:modeone}
\Psi^1 = i
\sqrt{\omega + k}\, \csc \quarterp\, \sech u \sqrt{\omega \cosh 2 u + k}\,
e^{i \alpha} \left[ e^{+i \chi} \Gamma_0 + e^{-i \chi} \Gamma_\phi
\right] {\cal P} U\,,
\end{equation}
where we have introduced a convenient but otherwise arbitrary
overall factor,
$U$ is an arbitrary ($k$-dependent)
complex Weyl spinor, ${\cal P}$ is the projection operator
defined
in~(\ref{eq:pdef}) and
\begin{equation}
\omega = + \sqrt{k^2 + 1}.
\end{equation}
Then from~(\ref{eq:useful}) we find
that the second spinor is determined to be
\begin{equation}
\label{eq:modetwo}
\Psi^2 =
\sqrt{\omega - k}\, \sec \quarterp
\, \sech u \sqrt{\omega \cosh 2 u - k}
\, e^{i \beta}\, \Gamma_* \Gamma_\phi
\left[ e^{+ i \tilde{\chi}} \Gamma_0 + e^{-i \tilde{\chi}} \Gamma_\phi
\right] {\cal P} U\,.
\end{equation}
In these expressions $\chi$ and $\tilde{\chi}$ are as defined
in~\cite{Minahan:2007gf},
\begin{eqnarray}
\label{eq:chidef}
e^{i \chi} &=&
\left(\frac{\sinh u + i \cos \halfp}{\sinh u - i \cos \halfp}\right)^{1/4}
(\tanh u + i \sech u)^{1/2}\,\cr
e^{i \tilde{\chi}} &=&
\left(\frac{\sinh u - i \cos \halfp}{\sinh u + i \cos \halfp}\right)^{1/4}
(\tanh u + i \sech u)^{1/2}\,,
\end{eqnarray}
while $\alpha$ and $\beta$ are given by
\begin{eqnarray}
e^{i \alpha} &=& e^{ - i \omega v + i k u}
e^{\frac{i}{2} \left( +\tan^{-1} (\omega \sinh 2 u) -
\tan^{-1}(k \tanh 2 u) \right) } \cr
&=& e^{-i \omega v + i k u}
\left( \frac{1 + i \omega \sinh 2 u}{1 - i \omega \sinh 2 u}
\frac{1 - i k \tanh 2 u}{1 + ik \tanh 2 u}\right)^{1/4}\,,\cr
e^{i \beta} &=& e^{ - i \omega v + i k u}
e^{\frac{i}{2} \left( - \tan^{-1} (\omega \sinh 2 u) -
\tan^{-1}(k \tanh 2 u) \right) } \cr
&=& e^{-i \omega v + i k u}
\left( \frac{1 - i \omega \sinh 2 u}{1 + i \omega \sinh 2 u}
\frac{1 - i k \tanh 2 u}{1 + ik \tanh 2 u}\right)^{1/4}\,.
\end{eqnarray}

Finally we come to the Majorana condition $(\Psi^I)^* = \Psi^I$.
There are 16 complex linearly independent solutions to
the equations of motion~(\ref{eq:fermioneom}): eight positive
frequency solutions shown in~(\ref{eq:modeone}) and~(\ref{eq:modetwo})
and eight corresponding negative frequency solutions given by
their complex conjugates.
After imposing the Majorana condition only eight complex
linearly independent solutions remain.
Upon quantization
the coefficient $U$ becomes an operator and
the Majorana condition relates
the coefficient
of positive frequency modes to the adjoint
of the coefficient of negative frequency modes, resulting in
a total spectrum of eight different kinds of fermionic particles.

It is also easy to exhibit manifestly real Majorana solutions
by starting with the particular linear combination
\begin{equation}
\label{eq:psione}
\Psi^1 = i\sqrt{\omega+k}\, \csc \quarterp
\, \sech u \sqrt{\omega \cosh 2 u + k}
\sum_\pm (e^{\pm i \chi} \Gamma_0 + e^{\mp i \chi} \Gamma_\phi)
(e^{+ i \alpha} U_\pm^1 + e^{-i \alpha} U^2_\pm)
\end{equation}
of positive and negative frequency modes.
In a Majorana representation of the $\Gamma$ matrices, in which
$(\Gamma^\mu)^* = - \Gamma^\mu$, $\Gamma^0$ being antisymmetric and
the rest symmetric, imposing the Majorana condition $(\Psi^1)^* = \Psi^1$
on~(\ref{eq:psione}) amounts to imposing
\begin{equation}
(U_+^1)^* = U_-^2\,, \qquad (U_+^2)^* = U_-^1\,.
\end{equation}
It is then not hard to see that $\Psi^1$ depends only on the
real parts of $U_+^1$ and $U_+^2$, which we will call $U^{1,2}$ respectively.
We conclude by displaying the resulting manifestly real solutions
\begin{eqnarray}
\label{eq:modethree}
\Psi^1 &=& 2i  \sqrt{\omega + k}\,\csc\quarterp\,
 \sech u \sqrt{\omega \cosh 2 u + k}
 \cr
&&\times
\left[ \Gamma_0 ( \cos \chi + \sin \chi\, \Gamma_{\phi \theta})
+ \Gamma_\phi ( \cos \chi - \sin \chi\, \Gamma_{\phi \theta})
\right]
\cr
&&\times
\left[ ( \cos \alpha + \sin \alpha \,\Gamma_{\phi \theta})
U^1 + (\cos \alpha - \sin \alpha \,\Gamma_{\phi \theta}) U^2 \right],
\end{eqnarray}
and
\begin{eqnarray}
\label{eq:modefour}
\Psi^2 &=& 2 \sqrt{\omega - k}\,\sec\quarterp\,
\sech u \sqrt{\omega \cosh 2 u - k}
\, \Gamma_* \Gamma_\phi  \cr
&&\qquad \times \left[ \Gamma_0 ( \cos \tilde{\chi} + \sin\tilde{\chi}\,
\Gamma_{\phi
\theta}) + \Gamma_\phi (\cos \tilde{\chi}- \sin \tilde{\chi}
\,\Gamma_{\phi \theta}) \right]
\cr
&&\qquad \times
\left[ (\cos \beta + \sin \beta \Gamma_{\phi \theta})
U^1 + (\cos \beta - \sin \beta \Gamma_{\phi \theta}) U^2\right],
\end{eqnarray}
which have been written in a form similar
to (2.25) and (2.28) of~\cite{Minahan:2007gf}.

\section{The 1-loop Functional Determinant}

We now apply the results of the previous two sections to check
the leading quantum correction to the energy of the giant magnon,
following well-established techniques for the semiclassical quantization
of solitons
(see in particular~\cite{Dashen:1975hd,Gervais:1976wr,Gervais:1975yg},
and~\cite{Jevicki:1979nr} for a review).
Of course
if energy is defined
as the Noether charge of
the actions~(\ref{eq:bosonicaction}) and~(\ref{eq:fermionaction})
associated with worldsheet $t$ translations then
the energy of any physical string state
is identically zero due to the
Virasoro constraints.
Rather than this kind of energy
we are instead interested in the conserved quantity
\begin{equation}
\Delta - J = \frac{\sqrt{\lambda}}{2 \pi} \int_{-\infty}^{+\infty}
dx \left[
(Y^0 \partial_t Y^5 - Y^5 \partial_t Y^0)
- (X^5 \partial_t X^6 - X^6 \partial_t X^5)
\right],
\end{equation}
where $\Delta$ is the charge associated with global time translations
in $AdS{}_5$ (i.e., the spacetime energy)
and $J$ is the U(1) charge associated with rotations in
the equator of the $S^5$.
The quantity $\Delta - J$
can be identified with the hamiltonian of physical
string excitations~\cite{Frolov:2002av}.  In light-cone gauge this
would be the familiar transverse hamiltonian.  In conformal gauge
we have to supplement the sigma-model
action with ghosts to cancel two unphysical bosonic degrees of freedom.
As discussed in section 2, it is sufficient for the purpose of our
semiclassical analysis to instead
simply omit the two massless
bosonic modes we found above.

The classical value of $\Delta - J$ for the
giant magnon solution~(\ref{eq:giantmagnon}) is
\begin{equation}
\label{eq:classicalenergy}
\Delta - J = \frac{\sqrt{\lambda}}{\pi} \left| \sin \frac{p}{2} \right|.
\end{equation}
The one-loop semiclassical correction to this result comes from
evaluating the functional determinant $\ln \det|\delta^2 S|$ around
the classical configuration.  The reader is
surely familiar with the fact that the contribution from this determinant is
\begin{equation}
\label{eq:oneloop}
\frac{1}{2} \sum_k (-1)^F \nu_k
\end{equation}
where $(-1)^F$ is $+1$ for bosons and $-1$ for fermions, and
$\nu_k$ are the frequencies of small oscillations around
the classical background.
This is manifestly independent of $\sqrt{\lambda}$ and hence
constitutes the ${\cal O}(1/\sqrt{\lambda})$ correction
to the classical result~(\ref{eq:classicalenergy}).

For a soliton at rest the $\nu_n$ are simply given by
the masses of the fluctuations.  The giant magnon moves
with velocity
\begin{equation}
{\mathrm v} = \cos \halfp
\end{equation}
and is only at rest for the special
case $p = \pi$.
The calculation of~(\ref{eq:oneloop}) in this case was performed
in~\cite{Minahan:2006bd}, where it followed as a special case of a
study of various more general spinning string configurations.
In fact the authors of~\cite{Minahan:2006bd} found that the first
${\cal O}(1/\sqrt{\lambda})$ correction to~(\ref{eq:classicalenergy})
vanishes, in agreement with the expectation based on
supersymmetry~\cite{Beisert:2005tm,Hofman:2006xt}
that the exact result for $p=\pi$ should be
\begin{equation}
\Delta - J = \sqrt{1 + \frac{\lambda}{\pi^2}} = \frac{\sqrt{\lambda}}{2}\left[
1 + \frac{0}{\sqrt{\lambda}} + {\cal O}(1/\lambda)\right].
\end{equation}

For general $p$ the expected relation~\cite{Beisert:2005tm,Hofman:2006xt} is
\begin{equation}
\Delta - J = \sqrt{1 + \frac{\lambda}{\pi^2} \sin^2 \frac{p}{2}}\,,
\end{equation}
which also has a vanishing ${\cal O}(1/\sqrt{\lambda})$ correction
to the classical value~(\ref{eq:classicalenergy}).
Now for general $p$ the magnon is no longer static so we employ the method
of Dashen, Hasslacher and Neveu~\cite{Dashen:1975hd} to
calculate the so-called stability angles $\nu_n$ which enter
in the one-loop functional determinant~(\ref{eq:oneloop}).
We begin by putting the system in a box of length $L \gg 1$, identifying
$x \cong x + L$.  From~(\ref{eq:giantmagnon}) it is clear that
the system is also periodic in time with period $T = L/{\mathrm v}$.
The stability angle $\nu$ of a generic fluctuation $\delta \phi$ is defined
as
\begin{equation}
\delta \phi(t + T, x) = e^{-i \nu} \delta \phi(t,x)\,.
\end{equation}

It is straightforward to read off the stability angles for the
fluctuations around the giant magnon from
our results in sections 2 and 3.
The four free massive $AdS_5$ bosons behave as
$e^{i k u - i \omega v}$ and hence trivially have
\begin{equation}
\nu_k(\delta Y) = {L \over {\mathrm v}} \frac{ \omega + {\mathrm v} k}{\sqrt{1 - {\mathrm v}^2}}\,.
\end{equation}
For the four physical $S^5$ fluctuations we find
from~(\ref{eq:spherefluctsolution})
that
\begin{equation}
\nu_k(\delta X)
 = {L \over {\mathrm v}} \frac{ \omega + {\mathrm v} k}{\sqrt{1 - {\mathrm v}^2}} + 2 \cot^{-1} k\,.
\end{equation}
Finally from the explicit form of the eight fermions
in section 4 we find
\begin{equation}
\nu_k(\vartheta)
= {L \over {\mathrm v}} \frac{ \omega + {\mathrm v} k}{\sqrt{1 - {\mathrm v}^2}} + \cot^{-1} k\,.
\end{equation}
Summing these results, with a minus sign for the fermions, gives
precisely the expected result that the
one-loop correction to~(\ref{eq:classicalenergy})
vanishes (even before integrating over $k$), thereby providing
a check on our results.

\acknowledgments{We are grateful to A.~Jevicki and A.~Volovich for several
helpful comments and to J.~Minahan for correspondence.
The research of MS is supported by NSF grant PHY-0638520.}

\end{document}